\documentclass[prb,twocolumn,showpacs,preprintnumbers,floatfix]{revtex4-1}

\usepackage{graphicx}
\usepackage{dcolumn}
\usepackage{bm}
\usepackage{latexsym,epsfig}
\usepackage{graphicx}
\usepackage{verbatim}
\usepackage{comment}
\usepackage{amsmath}
\usepackage{amssymb}
\usepackage{stmaryrd}
\usepackage{color}

\newcommand{\beq}{\begin{equation}}
\newcommand{\eeq}{\end{equation}}
\newcommand{\bea}{\begin{eqnarray}}
\newcommand{\eea}{\end{eqnarray}}
\newcommand{\ben}{\begin{eqnarray*}}
\newcommand{\een}{\end{eqnarray*}}
\newcommand{\bfig}{\begin{figure}}
\newcommand{\efig}{\end{figure}}

\newcommand{\ra}{\rangle}

\newcommand{\upa}{\uparrow}
\newcommand{\dna}{\downarrow}

\newcommand{\dg}{{\dagger}}
\newcommand{\pdg}{{\phantom\dagger}}

\newcommand{\nn}{\nonumber}

\begin{document}

\title{Quantum phases and phase transitions of frustrated hard-core bosons on a triangular ladder}

\author{Tapan Mishra$^1$, Ramesh V. Pai$^2$, Subroto Mukerjee$^{3,4}$,
and Arun Paramekanti$^{5,6}$}
\affiliation{$^1$International Center for Theoretical Sciences, Bangalore, 560 012, India}
\affiliation{$^2$ Department of Physics, Goa University, Taleigao Plateau, Goa 403 206, India}
\affiliation{$^3$ Department of Physics, Indian Institute of Science,
Bangalore, 560 012, India}
\affiliation{$^4$ Centre for Quantum Information and Quantum Computing (CQIQC), Indian Institute of Science,
Bangalore, 560 012, India}
\affiliation{$^5$ Department of Physics, University of Toronto, Toronto, Ontario, Canada M5S 1A7}
\affiliation{$^6$Canadian Institute for Advanced Research, Toronto, Ontario, M5G 1Z8, Canada}

\date{\today}

\begin{abstract}
Kinetically frustrated bosons at half-filling in the presence of a competing nearest neighbor repulsion
support a wide supersolid regime on the two-dimensional triangular lattice. We study
this model on a two-leg ladder using the finite-size density-matrix renormalization group method,
obtaining a phase diagram which contains three phases: a uniform superfluid (SF), an insulating
charge density wave (CDW) crystal and a bond ordered insulator (BO). We show that the transitions
from SF to CDW and SF to BO are continuous in nature, with critical exponents varying continuously
along the phase boundaries, while the transition from CDW to BO is found to be first order. The
phase diagram is also found to contain an exactly solvable Majumdar Ghosh point, and re-entrant
SF to CDW phase transitions.
\end{abstract}


\maketitle

\section{INTRODUCTION}
\label{sect:intro}

Quantum phase transitions in low dimensional systems are of great interest because quantum fluctuations play a greater role in their physics than in their higher dimensional counterparts. \cite{giamarchi_book,sachdev_book_01} For instance in one dimensional systems, quantum fluctuations can inhibit the appearance of long range order that is seen in higher dimensional systems with the same symmetries. An example of this is the one dimensional spin 1/2 Heisenberg chain with nearest-neighbor antiferromagnetic coupling which does not display any long range Ne\'{e}l order unlike its two dimensional square lattice counterpart. While quantum fluctuations in high dimensions maybe weak, geometrical frustration could prevent the occurrence of long range order. A naive expectation might be that quantum fluctuations and geometrical frustration, when present together would reinforce each other and severely inhibit ordering. However, there are several examples where the combination of frustration and quantum fluctuations induces interesting types of order such as through order-by-disorder transitions.~\cite{diep_book} In one dimension, where quantum fluctuations are especially strong, their interplay with frustration can give rise to interesting phases and phase transitions such as the recently proposed chiral Mott insulator in frustrated Bose-Hubbard ladders, a phase with a gap to all excitations
and a staggered pattern of equilibrium currents.~\cite{dhar_pra_12,dhar_arxiv_12}

In this paper, we study a frustrated one-dimensional system of hardcore bosons. In this model a boson on a site can hop to a neighboring site (with amplitude $t$) and also feels a density-density interaction (of strength $V$) from occupied neighboring sites. In addition to these terms, there is also a frustrating next nearest neighbor hopping term (of amplitude $t'$).  The Hamiltonian that describes this $t$-$t'$-$V$ model
is
\begin{eqnarray}
\!\!\!\!\!\! H \! &=& \!  - t \sum_i (a_{i}^{\dagger}a_{i+1}^{\phantom \dagger} + \text{h.c.})
 \! - \! t' \sum_i (a_{i}^{\dagger}a_{i+2}^{\phantom \dagger}+\text{h.c.}) \notag \\
 &+& \sum_i V n_i n_{i+1}
\label{eq:ham}
\end{eqnarray}
where $a_i^{\dagger}$ and $a_i^{\phantom \dagger}$ are creation and annihilation operators
for hard core bosons
at site $i$, and $n_i=a_i^{\dagger}a_i^{\phantom \dagger}$ is the boson number operator
at site $i$.
Here we have the constraint that $a_i^{{\dagger}2}=a_i^{2 \phantom \dagger}=0$, which avoids multiple
occupancies of the lattice sites. Frustration in this model arises from taking $t > 0$ and $t' < 0$. Further $V>0$ so that the interaction is repulsive. The model can be thought of as a zig-zag ladder of the sort shown in Fig.~\ref{fig:ladder} with the nearest-neighbor and next-nearest neighbor hopping arising from the motion of the bosons between and along the legs of the ladder respectively. Note that this model does not have a simple representation in terms of spinless fermions due to the presence of the next-nearest-neighbour hopping term, which when expressed in terms of spinless fermions will not correspond to a simple hopping term. Thus, our model even with $V=0$ is non-trivial and does not have a ground state corresponding to a filled Fermi sea. In fact, as we will see, for $V=0$, aside from the trivial point $t'=0$, there is only one other point corresponding to $t'=-t/2$, where the ground state can be obtained exactly.

\bfig[!t]
  \centering
  \includegraphics*[width=0.45\textwidth,draft=false]{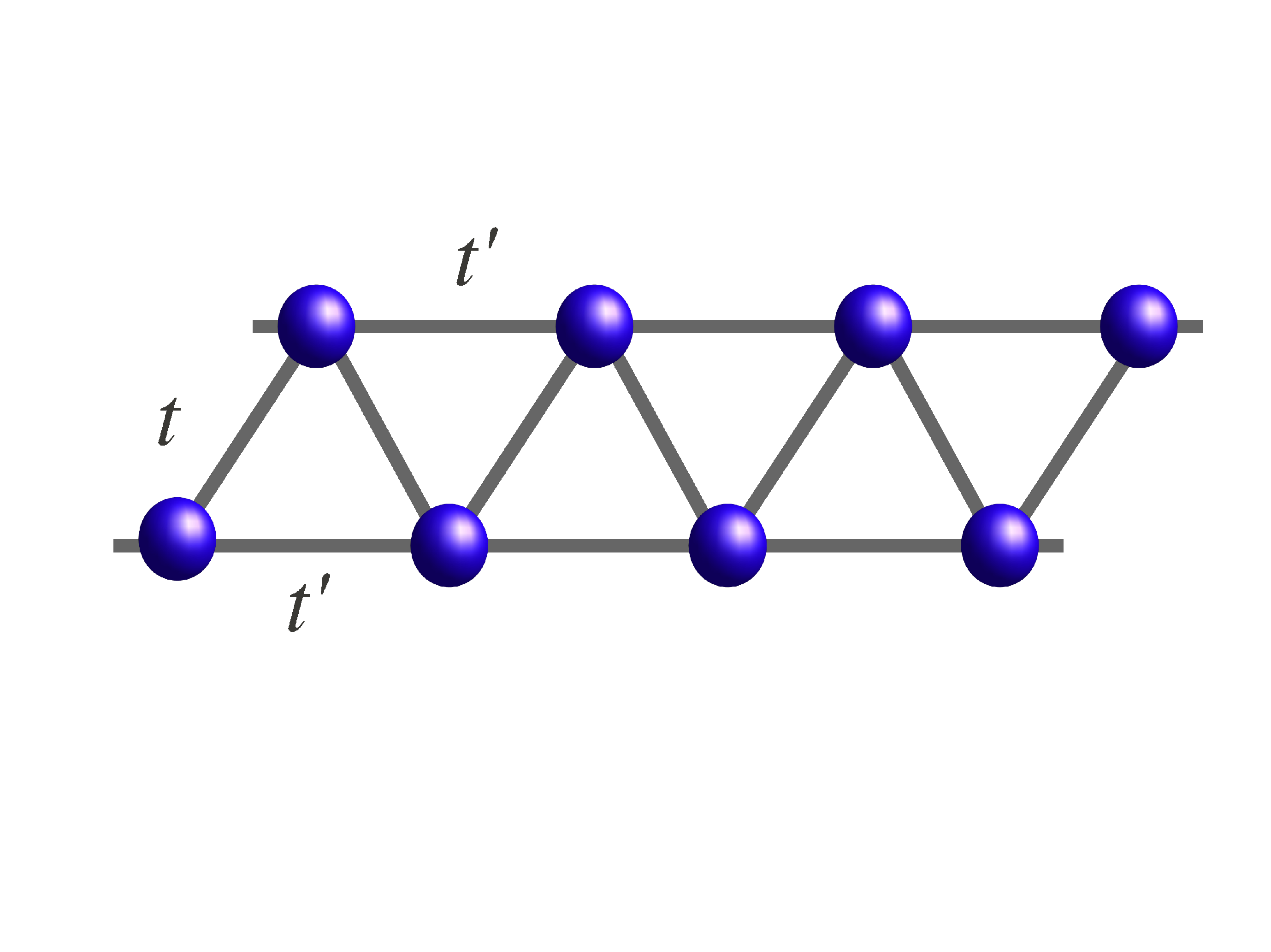}
    \caption{(Color online) Zig-zag ladder system representing the $t-t'-V$ model in one dimension. The
arrows are the representation of the hopping directions. $t>0$ and $t'<0$.}
    \label{fig:ladder}
\efig
Eq.~\ref{eq:ham} can be mapped onto a spin 1/2 Hamiltonian
by identifying $S^+_i = a_i^\dagger$,
$S^-_i = a_i^{\vphantom\dagger}$, and $S_i^z = (n_i-1/2)$.
Under this transformation the Hamiltonian \eqref{eq:ham} takes the form
\begin{eqnarray}
H = \sum_{i}\bigg[-2t(S_i^x S_{i+1}^x+S_i^yS_{i+1}^y)
+V S_i^zS_{i+1}^z \nonumber\\
- 2t'(S_i^xS_{i+2}^x+S_i^yS_{i+2}^y) \bigg]
\label{eq:hei}
\end{eqnarray}

By making a spin rotation on one sublattice, to change the sign of the nearest neighbor exchange coupling,
Eq.~\ref{eq:hei} can be recast as an anisotropic (XXZ type)
spin-$\frac{1}{2}$ model with a next-nearest-neighbor coupling. This model may then be viewed as a
specific easy-plane deformation of the well-known SU(2) symmetric $J_1$-$J_2$ model which has been
studied extensively.\cite{tonegawa_harada_87,burshil_farnell_95,chitra_ramasesha_95,white_affleck_96}. The
SU(2) symmetric $J_1$-$J_2$ model exhibits
gapless to gapped transition at $J_2 \approx 0.241167$.\cite{eggert_96} Interestingly, when $J_2= J_1/2$, the quantum system dimerizes in
the presence of frustration and is described by the Majumdar-Ghosh (MG) model \cite{majumdar_ghosh_69}. Variants of the
$J_1$-$J_2$ model with spin anisotropy have also been studied. The earliest such study was by Haldane who added an anisotropy parameter to the nearest-neighbor coupling term while the next-nearest-neighbor term was left isotropic.~\cite{haldane_82} More recently, a variant of the $J_1$-$J_2$ model with the same anisotropy in both the nearest-neighbor and next-nearest-neighbor terms has been studied.~\cite{furusaki_10,furusaki_12} Our model given by Eq.~\ref{eq:hei}, while anisotropic in spin is different from both the anisotropic models mentioned above in that, it has only a single $SU(2)$ symmetric point in parameter space (corresponding to $V=2t$ and $t'=0$) while the other two models have lines of points with $SU(2)$ symmetry. Some of the salient features of our model like continuously varying critical exponents along the phase boundary and re-entrant transitions are presumably due to the absence of such $SU(2)$ symmetric lines, which will be discussed later. We will show that our model also has a point in parameter space analogous to the MG model when $V=0$ and $t'=t/2$, but without $SU(2)$ symmetry, where we can obtain the ground state exactly. The model at this point can be thought of as a $U(1)$ analog of the MG model. Interest in this model also stems from the fact that the two-dimensional triangular lattice with such frustrated hard-core bosons has been shown to exhibit $\sqrt{3}\times \sqrt{3}$ supersolid phases~\cite{balents,vishwanath,heidarian}.

In this paper, we study the model given by Eqs.~\ref{eq:ham} at half filling numerically using the Density Matrix Renormalization Group (DMRG) algorithm in the entire parameter space of $t'<0$ and $V>0$ with $t>0$. We find that there are three phases, a superfluid (SF), bond-ordered (BO) phase and a charge density-wave (CDW). These are analogs of the gapless spin liquid, gapped dimerized phase and gapped Neel phase of the analogous anisotropic spin models. Our study mainly focusses on the phase boundaries and we show that the phase transitions in this model are of the Berezinski-Kosterlitz-Thouless (BKT) type and first order. Our results are  shown in the phase diagram of Fig.~\ref{fig:phasedia}. We find that along the lines of BKt transitions, there is a continuously varying critical exponent (corresponding to the Luttinger paramter). Another interesting feature of the phase diagram is a re-entrant SF-CDW-SF transition that exists in a part of parameter space and whose origin can be understood in terms of an instability of the CDW state.

\bfig[!t]
  \centering
  \includegraphics*[width=0.45\textwidth,draft=false]{phasedia2.eps}
    \caption{(Color online) Phase diagram of the half-filled $t$-$t'$-$V$ model. There exist two gapless to gapped
transitions, from SF to BO and SF to CDW for small $|t'|$ and $V$ values, which are of the BKT type. There is also a direct transition
from gapped BO to gapped CDW which is first order for large values of the interaction.
The point at $|t'|=0.5$ and $V=0$ is the Majumdar-Ghosh point and the dashed line is the theoretically calculated instability line for the CDW state. The hopping amplitude  is set to $t=1$. }
    \label{fig:phasedia}
\efig

The remainder of the paper is organized as follows. In section II, we study the MG point analytically and obtain the exact ground state to better understand the BO phase and in section III, we study the CDW phase and obtain an analytical form for the phase boundary between this phase and the SF highlighting the reentrant phase transition. In section IV, we describe the details of our DMRG study and give expressions for the various quantities that have been calculated and in section V, we present our numerical results for the full phase diagram along with a discussion of its different features.

\section{Majumdar-Ghosh point}

Consider the model $H \! = \!  - \sum_i (t~a_{i}^{\dagger}a_{i+1}^{\phantom \dagger}
 \! + \! t' a_{i}^{\dagger}a_{i+2}^{\phantom \dagger}+\text{h.c.})
$
for hard core bosons, where we have set the nearest neighbor repulsion $V=0$.
We can rewrite this Hamiltonian in the language of spin-1/2 operators, which
for $t'=- t/2$ is given by
\ben
H^{xy}_{MG} &=& \frac{t}{2} \sum_{k} \left[  -3/2 + \sum_{\alpha=x,y} (S^\alpha_k -S^\alpha_{k+1} +
S^\alpha_{k+2})^2 \right]
\een
Let us focus on a local state defined
on 3 successive sites $(j,j+1,j+2)$, of the
form $\vert \psi\ra = \vert (\upa_j \dna_{j+1} + \dna_j\upa_{j+1}) \sigma_{j+2} \ra$,
where $\sigma_{j+2}$ is an arbitrary spin state at site $j+2$.
It is easy to show that
\beq
(S^x_j -S^x_{j+1} + S^x_{j+2}) \vert \psi\ra =  \frac{1}{2} \vert (\upa_j \dna_{j+1} + \dna_j\upa_{j+1})
\bar\sigma_{j+2} \ra
\eeq
where $\bar\sigma_{j+2}$ denotes the original spin at site $j+1$ being flipped. This
means  $(S^x_j -S^x_{j+1} \! +\!  S^x_{j+2})^2 \vert \psi\ra =  \frac{1}{4} \vert \psi \ra$.
Similarly,
\beq
\!\! (S^y_j \! -\! S^y_{j+1} \!+\! S^y_{j+2}) \vert \psi\ra \!=\!  \frac{i \sigma_{j+2}}{2}
\vert (\upa_j \dna_{j+1} \!+\! \dna_j\upa_{j+1})
\bar\sigma_{j+2} \ra,
\eeq
so that
$(S^y_j -S^y_{j+1} \! +\!  S^y_{j+2})^2 \vert \psi\ra =  \frac{1}{4} \vert \psi \ra$. Thus,
the product state
\beq
\prod_{j\in {even}} \vert \upa_j \dna_{j+1} \!+\! \dna_j\upa_{j+1} \ra
\eeq
is an eigenstate of the Hamiltonian $H^{xy}_{MG}$, with energy per site $\frac{t}{2} ( -3/2+ 1/2)
= -t/2$. Since a sum of three spin-1/2 operators, such as $S^y_j -S^y_{j+1} \! +\!  S^y_{j+2}$, cannot
have a minimum eigenvalue of magnitude less than $1/2$, the state we have
found is evidently
a ground state. Another ground state, which is related by symmetry  to the above ground
state is simply
\beq
\prod_{j\in {odd}} \vert \upa_j \dna_{j+1} \!+\! \dna_j\upa_{j+1} \ra.
\eeq
While we have not proved that these are the only two
ground states of the Hamiltonian, our DMRG numerics indicate that this is the
case. Note that the ground states obtained above are not products of singlets like the usual MG state and are thus not spin rotation invariant. This is not surprising since our Hamiltonian is also not spin rotation invariant unlike the $J_1$-$J_2$ model.

\section{Phase boundary of the CDW state and the SF-CDW-SF re-entrant phase transition}
\label{sect:reentrant}

Consider the CDW state, which in spin language may be denoted as
$\vert \ldots\upa\dna\upa\dna\ldots \ra$. Let us define
Holstein-Primakoff bosons $h$, such that on the $\upa$-sites of the CDW, we have~\cite{holstein_primakoff_40}
\beq
S^z_i = (\frac{1}{2} -h^\dg_i h^\pdg_i)  ; S^+_i = h^\pdg_i; S^-_i = h^\dg_i
\eeq
while on the $\dna$ sites of the CDW, we have
\beq
S^z_i = (h^\dg_i h^\pdg_i-\frac{1}{2}) ; S^+_i = h^\dg_i; S^-_i = h^\pdg_i
\eeq
To quadratic order in the $h$-bosons, we find the Hamiltonian takes the form
\bea
H &=& - t \sum_i (h^\dg_i h^\dg_{i+1} + h^\pdg_i h^\pdg_{i+1}) + V \sum_i h^\dg_i h^\pdg_i \nn \\
&-& t' \sum_i (h^\dg_i h^\pdg_{i+2} + h^\dg_{i+2} h^\pdg_i)
\eea
Going to momentum space, and defining $\Psi^\dg_k = (h^\dg_k,h^\pdg_{-k})$,
we find the Hamiltonian
\bea
H = \sum_{k >0} \Psi^\dg_k
\begin{pmatrix} V \! - \! 2 t' \cos 2k & \! -2 t \cos k \\ \! -2 t \cos k & V \!-\! 2 t' \cos 2k \end{pmatrix}
\Psi_k,
\eea
with eigenenergy
$\lambda_k = \sqrt{(V-2t' \cos 2k)^2 - (2 t \cos k)^2}$. For large $V$, the spectrum has a
gap $\sim V$.
As $V$ decreases, however, the spectrum develops a soft mode signalling an instability
of the CDW state. For $|t'| < t/4$, the instability develops at $k=0$ and $k=\pi$ below a critical
repulsive interaction $V_{c,1} = 2(t-|t'|)$.
For $|t'| > t/4$, the instability develops at an
{\it incommensurate} wavevectors, $k_0$ and $\pi-k_0$, where
$k_0=\cos^{-1} (\frac{t}{4 |t'|})$, below a critical repulsive interaction
$V_{c,2} = 2 |t'| + \frac{t^2}{4 |t'|}$. The instability line is thus non-monotonic, with a
minimum at $|t'|=\frac{1}{2\sqrt{2}}$, and it approximately follows
the phase boundary of the CDW state found numerically using the DMRG. The non-monotonicity of the instability line
is responsible for the SF-CDW-SF re-entrant phase transition as can be seen from Fig.~\ref{fig:phasedia}.

\section{DMRG Treatment of the model}
We study the ground state of the model described by Eq.~\ref{eq:ham} using the finite-size DMRG method with open boundary
conditions.\cite{white_92,schollwock_review_05} This method is best
suited for (quasi-)one-dimensional problems.\cite{schollwock_review_05} For most of our calculations we study system sizes up to 300 sites and retain $128$ density
matrix eigen states with weight of the discarded states in the density matrix less than $10^{-6}$.
When $t'=0$, the model can be solved exactly using
the Bethe ansatz~\cite{cazalilla_citro_11} and there exists a
transition from a gapless SF to a gapped CDW phase like $(\ldots~1~0~1~0~1~0~1~0~\ldots)$
phase at $V=2t$. Here $1 \, (0)$ represents the presence(absence)
of a boson at a particular site. However, as we will show when the value of $t'$ is finite, the phase diagram is much richer.
In the present work we calculate various different physical quantities to characterize the phases and phase transitions of our model. The resulting phase diagram is shown in Fig.~\ref{fig:phasedia}.

We now list the various quantities we have calculated to identify the phases in our model. These quantities have been used previously to identify similar phases in
related models.~\cite{mishra_rigol_11} In order to distinguish between the gapped and gapless phases we calculate the single-particle excitation gap
\beq
G_L=E(L,N+1)+E(L,N-1)-2E(L,N).
\label{eq:gap}
\eeq
In Eq.~\eqref{eq:gap},
$E(L,N)$ is the ground-state energy of a system with $L$ sites and $N$ bosons.

The CDW phase and the transition into it can be studied by calculating the structure factor,
which is the Fourier transform of the density-density correlation function
\begin{equation}
 S(k)=\frac{1}{L^2}\sum_{i,j}{e^{ik(i-j)}(\langle{n_{i}n_{j}}\rangle-\langle n_i\rangle\langle n_j
\rangle)}.
\label{eq:str}
\end{equation}

The BO phase has a non-zero value of the bond-order parameter
\beq
O_{BO}=\frac{1}{L}\sum_i(-1)^i B_i,
\label{eq:obow}
\eeq
where
\beq
B_i=\langle a_i^\dagger a_{i+1}^{\phantom \dagger}+a_{i+1}^\dagger a_i^{\phantom \dagger}\rangle,
\eeq
and this is thus the quantity we calculate to identify thisphase and the phase transition into it.

The commensurate to incommensurate transition can be tracked by identifying the $k$ vector for ordering in a phase. This vector can be found by looking for a peak in the momentum distribution function
\beq
n(k)=\frac{1}{L}\sum_{i,j}{e^{ik(i-j)}\langle{a^{\dagger}_{i}a_{j}}\rangle}.
\label{eq:mom}
\eeq
Finally, to characterize the phase boundaries, we calculate the correlation function
\begin{equation}
 \Gamma(r)=\langle a_0^\dagger a_r\rangle,
 \label{eq:corr}
\end{equation}
which as we will see can be used to define the Luttinger parameter along the phase boundaries.
In the remainder of the paper, we set $t=1$.

\section{Results and discussion}
In the first part of this section we discuss the results of our calculation for $V=0$.
When $t'=0$, the system is a gapless SF  with finite momentum distribution at $k=0$. As $|t'|$ increases,
the system enters a gapped phase. The gapless to gapped transition can be seen by calculating the
single particle excitation gap as given in Eq.~\ref{eq:gap}. We plot the extrapolated gap $G_{L\rightarrow \infty}$ as a
function of $|t'|$ in Fig(~\ref{fig:gapther}).
\bfig[!t]
  \centering
  \includegraphics*[width=0.45\textwidth,draft=false]{gapther.eps}
    \caption{(Color online) The extrapolated gap $G_L \rightarrow \infty$ plotted as a function of $|t'|$ for $V=0$. The gapless
to gapped phase transition occurs at $|t'| \approx 0.33$}
    \label{fig:gapther}
\efig
The extrapolation is done using a third order polynomial in $1/L$. The transition point is located by observing where the extrapolated gap becomes of the order of
$10^{-5}$. In this way, we obtain the critical point for transition to the gapped phase at $|t'| \approx 0.33$. In Fig.~\ref{fig:gapinv} we show the polynomial extrapolation of the gap. It is clear
that the gap slowly becomes non-zero for values of $|t'| > 0.33$.
\bfig[!b]
  \centering
  \includegraphics*[width=0.45\textwidth,draft=false]{gapinv.eps}
    \caption{(Color online) Polynomial fitting of the gap shows that the gap slowly goes to zero for $|t'| \approx 0.33$. The the symbols are the numerical data and the lines are the fits.}
    \label{fig:gapinv}
\efig

The gap arises because of frustration which tries to make the ground state have bond order (BO) of the sort described in our analysis of the MG point. The bond energy is different for odd and even bonds in this phase. This emergence of the BO phase can be tracked by calculating the BO order parameter($O_{BO}$)
as given in Eq.~\ref{eq:obow}. In the BO phase, $O_{BO}$ is finite in the thermodynamic limit and vanishes for small values of $|t'|$.
We plot $O_{BO}$ as a function of $|t'|$ in Fig.~\ref{fig:obow} for three different lengths, $L=100, 200$ and $300$. The value of
$O_{BO}$ is maximum at $|t'|\approx 5.9$ and decreases on either side of the maximum. To the left of the maximum, i.e. when $|t'|$ decreases,
the jump becomes sharper and sharper as the length increases and appears to be saturating to a small value as $|t'|$ decreases.
\bfig[!b]
  \centering
  \includegraphics*[width=0.45\textwidth,draft=false]{obow1.eps}
    \caption{(Color online) $O_{BO}$ is plotted as a function of $|t'|$ for $L=100, 200$ and $300$. The increases of the value of $O_{BO}$
as $|t'|$ increases indicates the transition to the BO phase. The vertical dashed line indicates the location of the MG point and the horizontal dashed line, the value of $O_{BO}$ at this point, which is 0.5 independent of system size.}
    \label{fig:obow}
\efig
At $|t'| = 0.5$ the value of $O_{BO} = 0.5$ and independent of length as expected for the MG point.

Our analysis of sect.~\ref{sect:reentrant} shows that there is a commensurate to incommensurate transition at $|t'|=0.25$. We can verify this by calculating the momentum distribution $n(k)$ as a function of $k$ as shown in Fig.~\ref{fig:mom}. Here we see that the momentum distribution has one maximum at $k=0$ up to about $|t'| \approx 0.5$ after which the maxima shift to values of $k \neq 0$. This indicates the commensurate to incommensurate transition. The peak
position shifts towards $k= \pm \pi/2$ as $|t'|\rightarrow \infty$ where the two-leg ladder can be considered
as a single chain. The fact that the transition does not occur at $|t'|=0.25$ can be attributed to the hard core nature of the bosons. The analysis of sect.~\ref{sect:reentrant} assumed non-interacting particles.
 \bfig[!b]
  \centering
  \includegraphics*[width=0.45\textwidth,draft=false]{mom.eps}
    \caption{(Color online) Momentum Distribution $n(k)$ is plotted for different $|t'|$. There exists a commensurate to
incommensurate transition at $|t'| >  0.5 $ which can be seen as the onset of two peaks at $k\neq 0,\pi$.}
    \label{fig:mom}
\efig

Now we turn our attention to $V \neq 0$. When $|t'|=0$ it is known that at $V=2$, the model exhibits a transition from
SF to CDW phase~\cite{cazalilla_citro_11}. As the value of $|t'|$ is increased we get three different scenarios,\\

(1) SF - CDW transition as a function of $V$ for small values of $|t'|$\\

(2) BO - SF - CDW as a function of $V$ for intermediate values of $V$. \\

(3) BO - CDW transition as a function of $V$ for large values of $|t'|$\\

It is to be noted that the SF phase is gapless where as the BO and CDW phases are gapped. When the value of $|t'|$ is small,
the SF-CDW transition occurs at $V < 2$ since the presence of negative $|t'|$ suppresses the effect of $t$. This allows
the bosons to stabilize in the CDW ground state even for a small nearest neighbor interaction. This can be seen from
Fig.~\ref{fig:phasedia} where the upper phase boundary represents the SF - CDW transition for small values of $|t'|$. As $|t'|$ increases
the value of $V$ at which the transition takes place decreases gradually till it reaches a minimum and then increases again as expected from the analysis of sect.~\ref{sect:reentrant}. This indicates a re-entrant phase transition where at fixed $V$, an increase of $|t'|$ will drive the system from SF to CDW and back to SF again. It is interesting to note that the lowest point of the phase boundary obtained from the numerics is not too far from that predicted from the analytical calculation of sect.~\ref{sect:reentrant}.

The phase boundary between the SF and BO phases is also shown in Fig.~\ref{fig:phasedia}. This boundary originates at $V=0$ and $|t'| = 0.33$ and moves upwards as shown. The superfluid region is pinched off by the approach of the SF-BO phase boundary towards the SF-CDW boundary. Crudely speaking, $V$ prefers the formation of CDW order, $|t'|$ favors BO while the nearest neighbor hopping causes SF order. As long as both $V$ and $|t'|$ are smaller than or close to 1, the nearest neighbor hopping term ensures that the transition from BO to CDW has an intervening region of SF. However, once $V$ and $|t'|$ start becoming larger than 1, there is a direct transition from CDW to BO, which appears to be first order from our calculations.

We now discuss our characterization of these transitions. It is known that for $t'=0$, the transition from SF-CDW exhibits a BKT type scaling of the gap. The correlation function $\Gamma (r) \sim 1/r^\eta$ with $\eta =1$ at the critical point with an antiferromagnetic modulation and also a log correction. We find a similar BKT type scaling of the gap along both the SF-CDW phase boundaries and SF-BO phase boundaries. The scaling of the gap at the BKT transition can be used to locate the critical point of the transition fairly accurately.\cite{mishra_rigol_11}. The critical exponent $\eta$ can also be measured along the phase boundaries and we find this exponent to be varying continuously along both phase boundaries. We do not have the numerical accuracy to track this variation all the way up to the point where the boundaries appear to merge. It appears that there is a first phase transition from BO to CDW for larger values of $|t'|$ and $V$. In the following subsections, we present data supporting each of these claims.

\bfig[!b]
  \centering
  \includegraphics*[width=0.45\textwidth,draft=false]{gapscale_v0.eps}
    \caption{(Color online) The scaled gap $LG'_L$ is plotted as a function of $x_L$. (Inset) The scaled gap
plotted as a function of $|t'|$ for $V=0$. This shows the SF to BO transition at $|t'|=0.33$. }
    \label{fig:gaptv0}
\efig

\subsection{BKT scaling and gapless to gapped transition}

\label{subsect:gaplesstogapped}
Since the SF phase is gapless and the BO and CDW phase are gapped, we can use scaling of the gap $G_L$
given by Eq.~\ref{eq:gap} to locate the transition.
$G_L$ is computed for lattice up to $300$ sites. At the BKT transition from SF to BO the
gap closes as
\begin{eqnarray}
G\sim \exp\left[-\frac{a}{\sqrt{\big||t'|-|t'_c|\big|}}\right],
\label{eq:bkt}
\end{eqnarray}
where $a$ is a constant.


The correlation length $\xi$, which scales at the critical point as the inverse of the gap is
finite in the gapped phase and diverges in the gapless SF phase. We use the following
finite-size-scaling relation for the gap in the region close to the phase transition,
\begin{equation}
L G_L \times \left(1+\frac{1}{2\ln{L}+C} \right)= F\left( \frac{\xi}{L} \right),
\label{eq:scaling}
\end{equation}
where $F$ is a scaling function and $C$ is an unknown constant to be determined.
\bfig[!t]
  \centering
  \includegraphics*[width=0.45\textwidth,draft=false]{gapscale_v0pt5.eps}
    \caption{(Color online) The scaled gap $LG'_L$ is plotted as a function of $x_L$. (Inset) The scaled gap
plotted as a function of $|t'|$. This shows the SF to BO transition at $|t'|=0.32$. }
    \label{fig:gaptp}
\efig
In the region close to the critical point and within the SF phase, the values of $F(\xi/L)$ is expected to be
system-size independent, i.e.,
plots of $LG_L'=LG_L \left[1+1/\left(2\ln{L}+C\right) \right]$ as function
of $t'$ for different system sizes should intersect in that region. Also, the curves obtained by plotting
$L G_L'$ as function of $\xi/L$ for several values of $L$ should be system-size independent.
Therefore, the plots of $L G_L'$ as function of $x_L=\ln L -\ln \xi$ for different lengths collapse in the
critical region.
We obtain the values of $a$, $C$, and $|t'_c|$ for the best possible collapse
of the data in the gapped side where the correlation length diverges as
$\xi\sim \exp \left[ a/\sqrt{\big||t'|-|t'_c|\big|} \right]$. A similar procedure can
be used for the other part of the phase diagram i.e. the SF to CDW transition by replacing $|t'|$ by $V$.
The accuracy of this method has been tested by locating the Heisenberg point at $t'=0$.\cite{mishra_rigol_11}.
It is found to be at $V=2.02\pm 0.01$, very close to the analytical result.
Here, we also find the critical point for the SF to BO transition at $|t'| = 0.33\pm 0.01$ for $V=0$ which is
consistent with the value obtained previously in equivalent spin models.\cite{nomura_okamoto_94}

\bfig[!t]
  \centering
  \includegraphics*[width=0.45\textwidth,draft=false]{gapscale_tp.eps}
    \caption{(Color online) The scaled gap $LG'_L$ is plotted as a function of $x_L$. (Inset) The scaled gap
plotted as a function of $V$. This shows the SF to CDW transition at $V=1.31$. }
    \label{fig:gapv}
\efig
We use the above technique to obtain the boundaries between gapped and gapless phases in the phase diagram Fig.~\ref{fig:phasedia}. Another possible way to obtain the boundary is to extrapolate the gap to $L \rightarrow \infty$ and locate the points at which it goes from being non-zero to zero. However, we find that that the former technique is more accurate than the latter one and thus we use the BKT scaling form of the gap to locate the transition. In Fig.~\ref{fig:gaptp}
we show the scaling of the gap along the SF - BO transition boundary. The collapse of the
curves is obtained by plotting $L G_L'$ vs~$x_L$ within the gapped phase for $|t'_c| = 0.325\pm0.005$ and $V=0.5$(main panel).
The plots of $L G_L'$ vs~$|t'|$ (insets), for $V=0.5$ and for three values of $L$ show that the $LG_L'$ curves
intersect at the critical point $|t'|=0.32$. Similarly in Fig.~\ref{fig:gapv}, we show the collapse of the $L G_L'$ vs~$x_L$ data
for $|t'|=0.3$ when $V=1.31$ (main panel) and the intersection of the $LG_L'$ vs $V$ curves for different lengths at the
critical point (inset).

We now verify the locations of the SF-BO and SF-CDW phase boundaries by using the scaling of the BO order parameter and the
density-density structure factor. In the BO phase $O_{BO}$ is
finite and zero in the SF phase. The value of $O_{BO}$ is equal to $0.5$ at the MG point. In order to
see the the transition from BO to SF and then to CDW we start from the MG point i.e. from $|t'|=0.5$ and
then move along the $V$ axis. In Fig.~\ref{fig:vobow} we plot $O_{BO}$ as a fuction of $V$ for $L=100, 200, 300$. It can be seen
that the value of $O_{BO}$ decreases as we increase $V$. The decrease is faster for large lengths decrease implying that
$O_{BO}$ tends to zero in the thermodynamic limit. In order to see the actual transition point we perform a finite size
scaling of $O_{BO}$.\cite{ejima_nishimoto_07},\cite{mishra_rigol_11}
In Fig.~\ref{fig:linvobow}, we plot $O_{BO}$ as a function of $L^{-0.5}$ for different values of  $V$
 and then extrapolate to $L \rightarrow \infty$. It is evident that the curve for $V = 1.6$ extrapolates to
zero and the curves for $V > 1.6$ extrapolate to finite values showing the transition to BO phase for values of $V \approx 1.5$.
This result is in accordance with the phase diagram obtained by scaling of the gap where the transition point for the BO to SF transition
is at $V=1.53$.
\bfig[!t]
  \centering
  \includegraphics*[width=0.45\textwidth,draft=false]{obow2.eps}
    \caption{(Color online)$O_{BO}$ is plotted as a function of $V$ at $|t'|=0.5$ for different lengths.  }
    \label{fig:vobow}
\efig
\bfig[!t]
  \centering
  \includegraphics*[width=0.45\textwidth,draft=false]{obowscaling.eps}
    \caption{(Color online) $O_{BO}$ is plotted as a function of $1/L^{-0.5}$ for different $V$ at $|t'|=0.5$ showing the
BO-SF transition. }
    \label{fig:linvobow}
\efig
In order to understand the SF-CDW transition we perform a similar scaling of the density-density structure factor as given in
Eq.~\ref{eq:str}. An extrapolation shows that the $S(\pi)$ for $V \geq 1.6$ tends to a
finite value in the thermodynamic limit. However, for values of $V < 1.6$ the curves appear to extrapolate to zero. This is also
in accordance with the phase diagram where the SF-CDW transition occurs at $V=1.62$ for $|t'|=0.5$
\bfig[!t]
  \centering
  \includegraphics*[width=0.45\textwidth,draft=false]{scalespi.eps}
    \caption{(Color online)Finite size scaling of $S(\pi)$ shows the SF-CDW transition.   }
    \label{fig:linvspi}
\efig

\subsection{Critical exponent across the phase boundary}
We now compute the critical exponents across the two BKT phase boundaries representing the SF-CDW and SF-BO transition. This exponent $\eta$ is defined by the relation
\begin{equation}
 \Gamma(r) \sim 1/r^\eta
 \label{Eq:etadef}
\end{equation}
where $r=|i-j|$, $i$ and $j$ are the lattice indices. $\eta$ can thus be obtained by means of a straight line fit to $\Gamma(r)$ as a function of $r$ on a log-log scale. To avoid boundary effects, which can corrupt such a fit, we discard data obtained from the edges of our numerical system and use only data obtained from the bulk.
The exponent $\eta$ obtained this way will have a dependence on system size and the number of DMRG states kept in the calculation, which we have investigated. The plots of $\Gamma (r)$ vs. $r$ are shown in Fig.~\ref{fig:eta} for different points along the phase boundaries for $L=500$ and 128 states. About 50 sites from the bulk were used for the fit. The error bars mentioned are for the linear fits to the data. The values of $\eta$ while always close to 1, seem to be varying along parts of the phase boundaries. $\eta$ was found to decrease for all points with increasing $L$ and increasing number of DMRG states, so it is reasonable to believe that $\eta < 1$ along sections of the phase boundary. We find $\eta > 1$ for some parts of the phase boundary such as $t'=-0.3, V=1.3$ and $t'=-0.35, V=1.0$ and it is likely that the value of $\eta$ for these points will eventually go to 1 or lower with increasing system size and number of DMRG states which will have to be confirmed by more extensive numerical calculations. Further, the error in $\eta$ introduced due to the error in the determination of the critical points is minuscule and less than that due to the linear fits.

As can be seen from Fig.~\ref{fig:eta}$, \eta$ appears to be increasing as we move to larger values of $|t'|$ along the SF-CDW boundary. It is known that $\eta=1$ exactly at $t'=0$ but from our data it appears that $\eta$ drops as soon as a non-zero $t'$ is introduced and once again rises towards 1. Along the other phase boundary, $\eta$ starts from a value close to but less than 1 and increases as we move along it to larger values of $V$. It is likely that the two boundaries merge exactly at the point where $\eta=1$, which is also where the line of first-order transitions begins (a multicritical point). Another possibility is that the two phase boundaries merge before the line of first order points begins and there is an intermediate section along the phase boundary between CDW and BO, where $\eta$ varies continuously, i.e. the transition is Gaussian in nature. At the moment, we do not have the numerical accuracy to resolve these two scenarios.

\bfig[!t]
  \centering
  \includegraphics*[width=0.45\textwidth,draft=false]{eta.eps}
    \caption{(Color online)$\Gamma(r)$ plotted as a function of $r$ for $10 \leq r \leq 50$. The left panel is for the critical points
along the SF-CDW boundary and the right panel is along the SF-BO boundry of the phase diagram shown in Fig(\ref{fig:phasedia}). The symbols are
the value of $\Gamma(r)$ and the solid lines are the fitted function of the form $a/r^\eta$. The errors are obtained from the linear fit to the
data. It can be clearly seen that the value of $\eta$ increases by increasing the values of $V$ and $|t'|$ as we move along the SF-CDW and SF-BO
respectively. }
    \label{fig:eta}
\efig

It is interesting to note that we obtain a value of $\eta$ different from 1 and changing continuously at least along parts of the boundaries. This is not the case for previously studied anisotropic spin models. It has been argued that in those models, the value of $\eta$ is pinned to 1 along both the SF-CDW and SF-BO phase boundaries~\cite{haldane_82,furusaki_10,furusaki_12}. A renormalization group (RG) analysis of the sine-Gordon theory for these models shows that this is due to the fact that the zero umklapp line intersects a line of $SU(2)$ invariant points in the phase space of the models~\cite{haldane_82}. The point of intersection happens to be the transition point from spin fluid (SF in the language of hardcore bosons) to dimer order (BO in the language of hard core bosons) and has $\eta = 1$. The RG flow along both phase boundaries is towards this point thus pinning the exponent along them to 1. Further, in this RG analysis, the transition from CDW to BO is always continuous with a continuously varying $\eta > 1$, a Gaussian transition.

Our model does not have an $SU(2)$ symmetric point separating the SF and BO phases or a line of $SU(2)$ symmetric points in parameter space and thus will not have the same RG flow diagram. There is hence no reason {\em a priori} to expect the $\eta$ value to be pinned to 1 along the phase boundaries. However, since the transition out of the SF phase might still be expected to be governed by a sine-Gordon theory in which umklapp is not relevant, $\eta$ should be less than or equal to 1 along these boundaries. Thus, it can be expected that the values of $\eta > 1$ we seem to obtain for certain points along the phase boundaries will settle down to 1 or lower as the system size and number of DMRG states are increased.

An analytical understanding of the phase boundaries and critical exponents will require a detailed field theoretical study of the underlying sine-Gordon type action, which will be presented in a separate paper.

\subsection{Gapped to gapped phase transition}
\label{sect:gappedtogapped}
In this section we study the transition from BO to CDW phase at large $|t'|$ values.
For $|t'| > 0.6$ the SF phase shrinks very slowly and appears to finally disappear at $|t'| \approx 0.7$.
After this point the BO phase slowly undergoes a direct transition to the CDW phase.
To study the phase transition we use the CDW structure factor as an order parameter. We plot the extrapolated
values of $S(\pi)$ as a function of $V$ in Fig.~\ref{fig:vspi} for $|t'|=1.0$.
\bfig[!t]
  \centering
  \includegraphics*[width=0.45\textwidth,draft=false]{firstorder.eps}
    \caption{(Color online)$S(\pi)$ is plotted as a function of $V$ for $|t'|=1.0$ showing the BO-CDW transition.
(Inset)The $1/\xi_{L\rightarrow\infty}$ is plotted as a function of $V$ to confirm the first order nature of the BO-CDW transition. }
    \label{fig:vspi}
\efig
\bfig[!b]
  \centering
  \includegraphics*[width=0.45\textwidth,draft=false]{xiinv.eps}
    \caption{(Color online)Numerical data for $1/\xi_{L\rightarrow\infty}$ vs. $1/L$(symbols) and fits(lines) plotted for different values of $V$ to confirm the first order nature of the BO-CDW transition. }
    \label{fig:xiinv}
\efig
 The sudden jump in the value of
$S(\pi)$ at $V \approx 2.9$ implies a first order transition from the BO to the CDW phase. This transition is further
verified by examining the correlation length of the system. At the first order transition the single particle excitation
gap remains finite. As mentioned earlier, the correlation length $\xi \propto 1/G$. Therefore at the
first order transition $1/\xi$ should remain finite. We show the extrapolated values of $1/\xi_{L\rightarrow \infty}$ in the
inset of Fig.~\ref{fig:vspi}. It is evident from the figure that while approaching the transition
from the BO side, $1/\xi_{L\rightarrow \infty}$($G_L$) decreases rapidly, reaches a minimum at the transition point $V=2.97$
and then increases as the system enters the CDW phase. The extrapolation is done using a third order polynomial in
$1/L$ and is shown in Fig.~\ref{fig:xiinv}. The gap remains finite at the minimum implying the absence of an SF phase.
A further indicator of the first order nature of the transition comes from looking at the derivative of the ground state energy with
respect to $V$, $dE/dV$ shown in Fig.~\ref{fig:dedv} for different values of $|t'|$. It can be seen that there is the appearance at a
discontinuity in the derivative, which seems to get more pronounced with $|t'|$. This shows that the transition is discontinuous and the ``latent heat''
associated with it increases as one moves up the phase boundary.

\bfig[!t]
  \centering
  \includegraphics*[width=0.45\textwidth,draft=false]{dedv.eps}
    \caption{(Color online). The derivative of the ground state energy with respect to $V$ as a function of $V$. A discontinuity appears in the first derivative, showing that the ground state energy has a kink as a function of $V$, indicating a discontinuous transition. The ``latent heat'', proportional to the magnitude of the discontinuity seems to increase with increasing $|t'|$.}
    \label{fig:dedv}
\efig

The sine-Gordon theory for the CDW to BO transition predicts a continuous transition with $\eta > 1$ varying continuously along the phase boundary~\cite{haldane_82}.
However, it has been pointed out that this analysis ignores higher order umklapp terms, which can drive the transition first order, which is what we appear to be seeing here~\cite{furusaki_10}.
We emphasize, once again, that we do not have the resolution to clearly say if the first order line begins from the point of intersection of the SF-CDW and SF-BO phase boundaries.
There could be a small section of the phase boundary where the Gaussian transition between CDW and BO is seen. However, our numerics suggest that for sufficiently large values of $V$ along the CDW-BO phase boundary, the transition is of first order. We note that phase diagrams of the sort we have found can also be obtained from entanglement based studies of microscopic models such as in spin models on frustrated ladders.~\cite{kim}

\section{Conclusion}
\label{sect:conc}
We have presented a detailed study of the hardcore boson in a one dimensional lattice in the presence
of frustrated next-nearest neighbour hopping and the nearest neighbour interaction using the finite size DMRG method. The ground state phase diagram has three phases, SF, CDW and BO with continuous transitions along the SF-CDW and SF-BO phase boundaries and first order transitions along the CDW-BO phase boundary. The SF-CDW phase boundary is not monotonic giving rise to a re-entrant phase transition. Further from numerical data for our system sizes and density matrix states, the critical exponent $\eta$ appears to be different from 1 varying continuously along the SF-CDW and SF-BO phase boundaries in contrast to other anisotropic frustrated models.

\section{Acknowledgments}
SM thanks the Department of Science and Technology (DST) of the government of India for support. RVP thanks University Grants Commission (UGC) of the government of India for support.  AP was funded by NSERC of Canada, and acknowledges
support from the International Center for Theoretical Sciences (Bangalore) during the initial stages of this project. The authors would like to thank Diptiman Sen, Thierry Giamarchi and Luis Santos for very useful discussions.

\end{document}